\documentclass[12pt,preprint]{aastex}

\slugcomment{draft of \today}
\shorttitle{Propagation of UHE protons}
\shortauthors{Das {\it et al.~}}

\def\etal{{\it et al.}}
\def\eg{{\it e.g.,}}

\def\cm3{~{\rm cm^{-3}}}

\def\Mpc{~h^{-1}{\rm Mpc}}

\begin{document} 
\title{Propagation of Ultra-High-Energy Protons through the Magnetized Cosmic Web} 

\author{Santabrata Das\altaffilmark{1},
        Hyesung Kang\altaffilmark{2,3},
        Dongsu Ryu\altaffilmark{4},
    and Jungyeon Cho\altaffilmark{4}}

\altaffiltext{1}
{Astrophysical
Research Center for the Structure and Evolution of the Cosmos,
Sejong University, Seoul 143-747, South Korea: sbdas@canopus.cnu.ac.kr}
\altaffiltext{2}
{Department of Earth Sciences, Pusan National University, Pusan 609-735,
South Korea:\\ kang@uju.es.pusan.ac.kr} 
\altaffiltext{3}
{Author to whom correspondence should be addressed}
\altaffiltext{4}
{Department of Astronomy and Space Science, Chungnam National University,
Daejeon 305-764, South Korea: ryu@canopus.cnu.ac.kr, cho@canopus.cnu.ac.kr}

\begin{abstract}
If ultra-high-energy cosmic rays (UHECRs) originate from extragalactic
sources, understanding the propagation of charged particles through
the magnetized large scale structure (LSS) of the universe is crucial
in the search for the astrophysical accelerators.
Based on a novel model of the turbulence dynamo, we estimate the
intergalactic magnetic fields (IGMFs) in cosmological simulations
of the formation of the LSS.
Under the premise that the sources of UHECRs are strongly associated
with the LSS, we consider a model in which protons with $E \geq 10^{19}$
eV are injected by sources that represent active galactic nuclei 
located inside clusters of galaxies.
With the model IGMFs, we then follow the trajectories of the protons,
while taking into account the energy losses due to interactions with 
the cosmic background radiation.  
For observers located inside groups of galaxies like ours, 
about 70\% and 35\% of UHECR events above 60 EeV arrive within $\sim 15^{\circ}$
and $\sim 5^{\circ}$, respectively, of the source position 
with time delays of less than $\sim 10^7$ yr.
This implies that the arrival direction of super-GZK protons might
exhibit a correlation with the distribution of cosmological sources
on the sky.
In this model, nearby sources (within $10 - 20$ Mpc) should contribute 
significantly to the particle flux above $\sim10^{20}$ eV. 
\end{abstract}

\section{Introduction}

Over the past several decades,
significant progress has been made on both theoretical
and observational fronts in understanding the nature
and origin of ultrahigh energy cosmic rays (UHECRs), 
those with $E \ga 1$ EeV ($ = 10^{18}$ eV) 
\citep[for recent reviews, see][]{nagano00,bgg06}. 
Yet, the acceleration mechanism and the corresponding astrophysical 
``accelerators" of these energetic particles are still largely unknown.
Observational data from several experiments such as the High Resolution 
Fly's Eye (HiRes) indicate that the mass
composition of UHECRs becomes lighter at higher energies \citep{abbasi05}.
However, composition analyses that include high energy interactions are 
often model-dependent and inconclusive \citep{watson06}.
According to a recent report from the Pierre Auger Observatory,
the mass composition is likely mixed,
possibly becoming heavier above 30 EeV \citep{unger07}.
The overall distribution of UHECR arrival directions is 
considered to be consistent with isotropy \citep{bo03}.  
Exceptions to this general isotropy include the small-scale clusterings 
of doublets and triplets found in data collected by the Akeno Giant 
Air Shower Array (AGASA) experiment \citep{takeda99,uchihori00}.
In addition, a possible correlation of AGASA events with BL Lacertae objects
has been suggested \citep{tinyakov01}.
But these claims have been confirmed neither by HiRes \citep{abbasi06}
nor by Auger \citep{harari07,armen07}. 
However, the Auger Collaboration recently reported that
the arrival directions of UHECRs above $60$EeV in their data show
a correlation with the position of active galactic nuclei (AGNs)
lying within 75 Mpc \citep{augerScience}.

Since protons with $E \ga 1$ EeV cannot be confined within the
Galactic plane,  UHECRs likely originate from extragalactic sources.
In particular, the overall isotropy of arrival directions 
suggests that there may be a large number of sources distributed 
over cosmological distances \citep{nagano00,bo03}.
During their propagation through intergalactic space, such protons will lose
energy by means of pion and pair production processes while
interacting with the cosmic background radiation \citep{g66,zk66}. 
The flux of ultra-high-energy (UHE) protons from cosmological sources is thus
expected to be strongly attenuated,
resulting in a significant suppression in the observed spectrum above
the GZK threshold energy, $E_{\rm GZK}\approx 40$ EeV \citep{bgg06}.
Although the AGASA data show no indication of the GZK suppression
\citep{nagano00},
both the Yakutsk Extensive Air Shower array and HiRes have reported a suppression of flux above
$E_{\rm GZK}$, contradicting the AGASA finding \citep{pravdin99,zech04}.
Indeed the same suppression was seen in a recent Auger measurement 
\citep{facal07}, which seems to have ended the 
controversy over the presence of the GZK cutoff.
The so-called dip-calibrated UHECR spectra from different
experiments compiled by \citet{bgg06},  
appear to be in good agreement with each other 
and to be consistent with GZK suppression.
However, it has yet to be understood whether this suppression is actually due to the
GZK cutoff or due to the maximum acceleration energy,
$E_{\rm max}$, of astrophysical accelerators.  
If UHECRs are protons,
the GZK energy loss should operate at acceleration sites as well,
leading to an $E_{\rm max}$ close to $E_{\rm GZK}$ \citep[see, \eg][]{krb97}.

Most UHE protons observed above the GZK energy must come from
within the so-called GZK sphere of radius $R_{\rm GZK} \sim 100 $ Mpc, although
the proton interaction length at $E_{\rm GZK}$ is $l_{\rm 40EeV}
\approx 1$ Gpc, corresponding to $z\sim 0.2$  \citep{berezinsky88}.
However, finding cosmological sources inside the GZK sphere from the 
arrival directions of UHECRs is not straightforward, 
since their paths are deflected by intergalactic magnetic fields 
(IGMFs) \citep[e.g.][]{sigletal03,dolagetal04}.
Sigl and collaborators have extensively studied the propagation of UHECRs
in a structured and magnetized universe, adopting a numerical model for 
the IGMFs \citep{sigletal03,sigletal04,armen05}.
In this model, the IGMFs are generated by means of the Biermann ``battery'' 
mechanism at shocks and then evolved passively in a cosmological hydrodynamic
simulation \citep{kulsrude97,rkb98}. 
The strength of the resulting fields is rescaled to match the
simulated field strength in Coma-like clusters to the observed
strength, which is on the order of microgauss.
UHECRs with $E\ge10$EeV are then injected at cosmological sources.
These particles propagate through the magnetized large scale structure
(LSS) of the universe, and arrive at a mock observer with deflection
angle $\theta$, the angle between the arrival direction and the
source's location on the sky. 
Sigl \etal\ found that the deflection due to IGMFs is significant,
with $\theta \ga 20^{\circ}$ above 100 EeV.  
On the other hand, Dolag \etal\ adopted an IGMF model from a 
``constrained'' cosmological simulation employing a magnetohydrodynamic (MHD)
version of a smoothed particle hydrodynamics (SPH) code. 
They found the deflection angle of protons with 100 EeV
to be less than $1^{\circ}$, contradicting the estimate by
Sigl \etal\ \citep{dolagetal04,dolagetal05}.

This controversy over the predicted deflection angles demonstrates 
the importance of modeling the IGMFs in identifying
the astrophysical sources and studying the origin of UHECRs.
In order to reexamine this issue, here we adopt IGMFs, 
based on a novel models of the turbulence dynamo \citet{ryuetal08} 
In this model, the strength of the magnetic fields is
estimated from the local vorticity and turbulent kinetic energy
in cosmological structure formation simulations.
For the field direction, the passive fields from these simulations are used.

The maximum energy of nuclei of charge $Z$ that can be confined and
accelerated by astrophysical sources is given by
\begin{equation}
E_{\rm max} \approx Z \left({V \over c}\right)
\left({B \over \mu{\rm G}}\right)
\left({L \over {\rm kpc}}\right) 10^3 {\rm EeV},
\end{equation}
where $V$, $B$, and $L$ are the characteristic flow speed, magnetic
field strength, and linear size of the accelerator, respectively \citep{hillas84}.
There are a few viable candidates that can produce the required
$E_{\rm max} \sim 100$ EeV: jets from AGNs \citep[\eg][]{biermann87}, 
gamma-ray bursts \citep[\eg][]{waxman95}, and cosmological shocks
\citep{krj96,krb97}.
In this study, we consider AGNs inside galaxy clusters as the sources
of UHECRs.
Thus, the source position is in effect correlated with the LSS of
the universe.
Protons with $E\ge 10$EeV are injected at the sources and travel
through the simulated magnetized space until they lose energy down to
10 EeV, meanwhile visiting mock observers placed inside groups of galaxies.
In this study we focus mainly on the deflection angle, the time delay
relative to rectilinear flight, and the energy spectrum of the UHE protons.

In the next section, we describe our model IGMFs, cosmic-ray sources, and
observers, and the simulations of the propagation of UHE protons
in intergalactic space. 
The results are presented in \S 3.
Finally, we conclude in \S 4. 

\section{Models and Methods}

\subsection{Intergalactic Magnetic Fields}

Our cosmological hydrodynamic simulations of a concordance $\Lambda$CDM 
universe are carried out with the following parameters:
$\Omega_{\rm BM}=0.043$, $\Omega_{\rm DM}=0.227$, and
$\Omega_{\Lambda}=0.73$, $h \equiv H_0$/(100 km/s/Mpc) = 0.7, and
$\sigma_8 = 0.8$.
A cubic box of comoving size $100 \Mpc$ is simulated using $512^3$
grid zones for gas and gravity and $256^3$ particles for dark matter. 
In the simulation, magnetic fields are generated through the Biermann battery
mechanism at structure formation shocks and then evolved passively with the
flow motions \citep{kulsrude97,rkb98}.
The simulation was repeated for six different realizations of the initial
conditions to examine the effects of cosmic variance. 
Non gravitational effects, including radiative cooling, 
photo ionization and heating, and feedback from star formation are ignored.
Those processes affect the generation and evolution of magnetic fields 
mainly on small scales \citep{krco07}, which should not alter the
large-scale fields primarily responsible for the deflection of UHECRs.  

In principle, if we were to perform full MHD simulations, we could follow
the growth of the IGMFs through stretching, twisting, and folding of
field lines, the process known as the turbulence dynamo.
In practice, however, the computational resources currently available 
do not allow high enough numerical resolution to reproduce the full 
development of MHD turbulence:
since the numerical resistivity is larger than the physical resistivity
by many orders of magnitude, the growth of the magnetic fields saturates
before the dynamo action becomes fully operative \citep{kulsrude97}.
So, \citet{sigletal03,sigletal04} rescaled 
the strength of their passively evolved fields 
in the postprocessing analysis 
to match the observed field strength in clusters of galaxies.
This rescaling, which hinges on the observed field strength in
the intracluster medium, does not necessarily result in correct
field strengths for filaments, sheets, and voids.
On the other hand, in the MHD SPH simulations of \citet{dolagetal04,
dolagetal05}, the initial field strength was adjusted to obtain
a microgauss level in clusters of galaxies at the present epoch. 
They demonstrated that their simulated cluster fields are consistent
with various observations, such as rotation measure profiles and
the total radio powers of cluster halos.
A lack of observations, however, prevents their simulated fields in 
low-density regions from being tested against the real magnetic fields
in filaments and sheets. 
In the MHD SPH simulations, the flow motions can be resolved
reasonably well in high-density regions, where the smoothing length is
sufficiently small, ensuring adequate growth of magnetic fields. 
However, turbulence may not be fully realized in the low-density
regions, which have large smoothing lengths, so the field strength in 
filaments and sheets may be underestimated in such simulations. 

In this study, we take a new approach detailed in \citet{ryuetal08}.
If one assumes that magnetic fields grow as a result of the turbulence dynamo, 
their energy density can be estimated from the eddy turnover number and
turbulent energy density as follows :
\begin{equation}
\varepsilon_B = \phi \left({t \over t_{\rm eddy}}\right)
\varepsilon_{\rm turb}.
\end{equation}
Here the eddy turnover time is defined as the reciprocal of the vorticity
at driving scales, $t_{\rm eddy} \equiv 1/\omega_{\rm driv}$ 
(${\vec \omega} \equiv {\vec\nabla}\times{\vec v}$), and $\phi$ is the
conversion factor from turbulent to magnetic energy, which
is determined from high-resolution MHD simulations of turbulence. 
For our model IGMFs, the number of eddy turnovers is estimated as
the age of the universe multiplied by the magnitude of the local vorticity,
that is, $t_{\rm age}\omega$.
The local vorticity and turbulent energy density are calculated
from the cosmological structure formation simulations.
The energy density given by equation (2) fixes the strength of the IGMFs,
so our model requires neither rescaling of the field strength
nor adjustment of the initial fields.
As in the work of Sigl \etal, we assume that the topology of the IGMFs in the LSS
can be represented statistically by the topology of the passive
magnetic fields in the cosmological simulations.  

Figure 1 shows a two-dimensional slice of the magnetic field strength
in our model at the present epoch.
The IGMFs are structured and well correlated with the weblike cosmic 
distribution of matter.
The strongest magnetic fields, with $B \ga 0.1\mu$G, are found inside
and around clusters, while the fields are weaker in filaments,
sheets, and voids.
Overall, there is a correlation between field strength and gas
density, as can be seen in Figure 2({\it left}).
In the regions of galaxy clusters, with 
$\rho_{\rm gas}/\langle \rho_{\rm gas} \rangle \ga 10^3$, 
we find $\langle B \rangle \sim 1~\mu$G.
For typical filamentary regions, with
$\rho_{\rm gas}/\langle \rho_{\rm gas} \rangle \sim 10$,
the field has $\langle B \rangle \sim 10^{-8}$ G.
By comparison, the average field strength in filaments is found to be 
$\langle B \rangle \sim 10^{-7}$ G by \citet{sigletal04} and
$\langle B \rangle \sim 10^{-10}$ G by \citet{dolagetal05}.
The right panel of Figure 2 shows the volume fraction, $df/d \log B$
({\it solid line}), and its cumulative distribution, $f(>B)$ ({\it dotted line}) and $f(<B)$
({\it dot-dashed line}).
In our model, the volume filling factor for $B >10^{-8}~{\rm G}$
is $f(>10^{-8}{\rm G}) \approx 0.01$.
By comparison, $f(>10^{-8}~{\rm G}) \approx 0.1$ in \citet{sigletal04}  
and $f(>10^{-8}~{\rm G}) \approx 10^{-4}$ in \citet{dolagetal05}.
Hence, the field strength in filaments in our model lies
in the middle of the values from the models used by these two groups,
that is, lower than Sigl \etal's but higher than that of Dolag \etal's.

\subsection{AGNs as UHECR Sources}

In the framework of bottom-up acceleration models, AGNs are the most
studied candidate astrophysical accelerators to produce
cosmic-ray nuclei beyond the GZK energy \citep[for a review, see ][]{bgg06}.
As noted in \S I, it has also been suggested that cosmological shocks
may accelerate nuclei of charge $Z$ up to an $E_{\rm max}$ of 
a few times $10^{19}\ Z$ eV, but it is unlikely that protons would be 
accelerated beyond the GZK energy by such shocks \citep{krb97, inoue07}.
However, if some UHECRs are iron nuclei, cosmological shocks can
provide the acceleration sites for super-GZK cosmic rays \citep{inoue07}.
In this paper, we consider only AGN-like objects as sources
of UHECRs, since we focus on the propagation of UHE protons.

We identify X-ray clusters characterized by the X-ray emission weighted gas temperature
$kT>1$ keV  in the cosmological simulations as source locations. 
One AGN is placed at the center of each ``host'' cluster.
With this selection criterion, the source locations are chosen, in effect,
at high-density regions with the strongest field strength.
Table 1 shows the number of such clusters found in the six cosmological
simulations ($\Lambda$CDM1-$\Lambda$CDM6) with different initial
conditions.
The three-dimensional distribution of the 18 sources in $\Lambda$CDM1
is shown in Figure 3.
Given the simulation volume of $(100 \Mpc)^3$, the mean separation of
sources is $l_s \approx 40 \Mpc$ and the source number density is
$n_s = 2-3 \times10^{-5} h^{3}{\rm Mpc}^{-3}$,
which is consistent with the required UHECR source density
inferred from the small-scale clustering found in the AGASA data
\citep{Yoshiguchietal03, sigletal03, bm04}.
The field strength at the source locations mostly lies in the range
$ 0.1 \mu {\rm G} \la B_s \la 2 \mu {\rm G}$, with a peak at $\sim 1 \mu$G;
its distribution is shown in Figure 4 ({\it left}).

\subsection{Groups of Galaxies as Mock Observers}

The key physical condition for an ``observer'' that is most relevant to
this study is the strength and direction of the magnetic fields, since we
are interested primarily in the deflection angles and time delays of
UHECRs. 
Little is known about the magnetic fields in the intergalactic space
within the Local Group.
So, we select groups of galaxies identified in the simulation data that
have similar halo gas temperatures to the Local Group,
that is, $ 0.05 {\rm keV} < kT < 0.5 {\rm keV}$ \citep{rasmus01},
assuming that these groups are located in magnetic environment    
similar to that of the Local Group.  
There are about $1000-1400$ identified groups with gas temperatures
in this range inside the simulation volume (see Table 1).
As can be seen in Figure 3, these groups are not distributed uniformly 
but are located mostly along filaments, following the matter distribution
of the LSS.
A mock observer, modeled as a sphere of radius $R_{\rm obs}=0.5 \Mpc$, 
is placed at each group.
The value of $R_{\rm obs}$ is chosen so that the observer's sphere
is well contained within the associated filament, since the typical
thickness of the magnetized region around a filament is about
$2-3 \Mpc$. 
If we were to use an $R_{\rm obs}$ smaller than the value adopted, 
smaller cross sections would lead to smaller detection rates of cosmic rays 
in our numerical experiment described below. 
The right panel of Figure 4 shows the field strength within the
observer spheres for the six cosmological simulations: 
$ 10^{-4}  {\mu \rm G} \la B_{\rm obs} \la 0.1 {\mu \rm G}$ with a peak at
$\sim 10^{-9}$ G.
This illustrates the distribution of magnetic field strength in filaments. 

\subsection{Propagation of UHE protons in the Intergalactic Space}

The characteristics of the particle spectrum accelerated at UHECR
sources are largely unknown.
We thus make the simple assumption that the accelerated protons have
a power-law energy spectrum at their source: $N_{\rm inj}(E_{\rm inj})
\propto E_{\rm inj}^{-\gamma}$ for 10 EeV $\le E_{\rm inj} \le 10^3$ EeV.  
In practice, $\gamma=0$ is used to generate a flat injection spectrum 
at the sources, and later a weighting factor proportional to
$E_{\rm inj}^{-\gamma}$ is applied to the statistics (except for the distribution
in Fig. 5; see \S3.1for details).
For each $\Lambda$CDM simulation,
a total of $3\times10^4$ particles are randomly distributed over the sources 
and then launched in random directions  
from random positions inside a sphere of radius $0.5\Mpc$. 

We follow the trajectories of the UHE protons by numerically integrating
the equations of motion in our model IGMFs, 
\begin{equation}
\frac{d\vec{r}}{dt}=\vec{v} ~ ; ~~~~ \frac{d\vec{v}}{dt}=\frac{Ze}{mc}
\left(\vec{v}\times \vec{B} \right).
\end{equation}
The energy losses due to photo pion production and pair production
are treated with the continuous loss approximation \citep{bgg06},
but the adiabatic losses due to cosmic expansion are ignored,
because the largest source-to-observer distance is $D_{\rm max} \sim 1$
Gpc in our experiment, corresponding to only  $z_{\rm max} \approx 0.2$.
In practice, the distances are not known in advance, since we are integrating
the trajectory from sources to observers.

The simulation box of $(100 \Mpc)^3$ at $z=0$ is used to
define sources (host clusters), mock observers (groups), and the
IGMF data.
Additional virtual boxes with the same distribution of mock
observers and IGMF data are periodically stacked indefinitely.
Particles are injected from the sources only in the original box.
Then they travel through the magnetized space consisting
of the original box and the replicated periodic boxes.  
Once a particle visits an observer sphere,
the arrival direction, time delay and energy of the particle are
registered as a ``recorded event.''
We let the particle continue its journey, visiting several observers
during its full flight, until its energy falls to 10 EeV.
With $3\times 10^4$ protons injected for each simulation box, about
$2.6 \times 10^5$ events are recorded in total for all six
$\Lambda$CDM simulations.

In our propagation experiment, the source-to-observer distance $D$
can be arbitrarily small because the specific way that we set up source
and observer locations, and the statistics of recorded events depend 
on the minimum value of $D$. 
In reality, the closest AGNs to us are Centaurus A, at 3.42 Mpc 
\citep{ferrarese07}, in the southern hemisphere and M87, at 16.7 Mpc
\citep{mei07}, in the northern hemisphere. 
So we mostly present the results for $D_{\rm min}=3$ Mpc or $D_{\rm min}=10$ Mpc.
The recorded events with $D < D_{\rm min}$ are excluded from the analysis.
For $D_{\rm min} \ge 10$ Mpc, however, the results become less sensitive
to the value of $D_{\rm min}$. 

\section{Results}

\subsection{Deflection Angle and Time Delay of UHECRs}

With a gyroradius
\begin{equation}
r_g = 10\ {\rm kpc}\left({E \over 10^{19}{\rm eV}}\right)
\left({B \over \mu{\rm G}}\right)^{-1},
\end{equation}
UHE protons will suffer significant deflection during their propagation
when they pass regions with $B \ga 10^{-8}$ G, that is, clusters and filaments.
Here the filamentary regions are more significant players than clusters, since
the volume filling factor of filaments is much larger than that of
clusters (see Fig. 2).
As a result of this deflection, the actual path traveled by the UHECRs in the 
presence of the IGMFs can be much longer than a rectilinear path, 
causing a significant time delay.
We therefore measure the deflection angle as the angle between the arrival
direction of the cosmic rays and the source position on the sky,
and the time delay as the difference between the arrival time and
the rectilinear travel time.

UHE protons can be also deflected inside the host clusters of sources,
before they escape to intergalactic space.
Typical clusters have a magnetized core envelope structure 
with $B_{\rm core} \sim 1\mu$G, $R_{\rm core} \la 0.5 \Mpc$ and
$B_{\rm env} \sim 0.01-0.1 \mu$G, extending out to $R_{\rm env}\sim
3 \Mpc$ (see Fig. 1).
So, the protons with $E \la E_{\rm GZK}$ injected by AGNs are 
scattered by turbulent magnetic fields inside the host clusters 
and confined within the magnetized structure for a while.
The scatterings by the turbulent fields local to the sources alone
can cause a deflection angle
\begin{equation}
\theta_{\rm source} \sim \tan^{-1}\left({{\rm a~few~Mpc} \over D}\right), 
\end{equation}
for protons with $E \la E_{\rm GZK}$.

Figure 5 shows the distributions of the deflection angle $\theta$
and the time delay $t_d$ as functions of the distance $D$ for the
$\Lambda$CDM1 simulation.
The events are divided into three channels in observed energy,
as follows: $10<E_{\rm obs}<30$ ({\it top}), $30<E_{\rm obs}<60$
({\it middle}), and $E_{\rm obs}>60$ ({\it bottom}), where the
particle energy is given in units of EeV.
The data points are color-coded by the injection energy in the same
three channels, that is, red for $10<E_{\rm inj}<30$, blue 
for $30<E_{\rm inj}<60$, and green for  $E_{\rm inj}>60$.
For these plots, the calculation was performed with an injection spectrum
with $\gamma=2.7$ instead of the flat spectrum, since the recorded
data points in this type of representation cannot be weighted with
a factor proportional to $E_{\rm inj}^{-\gamma}$.

Sub-GZK protons, with $E_{\rm obs}< 60$ EeV, come from sources
as distant as $D\sim 1$ Gpc. 
For these particles, the distribution of $\theta$ 
shows a pattern roughly in accord with the diffusive transport limit,
but it also indicates a bimodality divided at $D\sim 15$ Mpc 
(Fig. 5, {\it top and middle left}).
The events with $D \la 15$ Mpc are likely to be cases in which
both the source and observer belong to the same filament.
These particles are more likely to travel through strongly 
magnetized filaments rather than void regions, resulting in 
large deflection angles in addition to a large $\theta_{\rm source}$ 
given by equation (5).
On the other hand, for the events with $D \ga 15$ Mpc the particles
come from distant sources associated with different filaments. 
Some of these may fly through voids and arrive with small $\theta$,
while most are deflected significantly by the IGMFs.
On the other hand, on average the time delay tends to increase with
distance $D$, as expected.

Super-GZK protons, with $ E_{\rm obs}>60$ EeV come mostly from within $D\sim 100$ Mpc.
Most of them from $D \ga 15$ Mpc arrive with $\theta \la 10^{\circ}$
and $t_d \la 10^7$ yr (Fig. 5, {\it bottom}).  
Since the volume filling factor for $B $ greater than a few times $10^{-8}$G
(corresponding to $r_g \sim 2-3$ Mpc) is small, those particles could
travel almost rectilinearly through void regions, avoiding the
strongly magnetized regions of clusters and groups.

In Figure 6, the distribution of recorded events is shown in
the $({E}_{\rm obs}, \theta)$ and $({E}_{\rm obs},\log t_d$) planes.
The events recorded in all six $\Lambda$CDM simulations are included
and $\gamma=2.7$ and $D_{\rm min}=3$ Mpc are used.
On average the deflection angle decreases with energy, but
a clear transition from the diffusive transport regime to the
rectilinear propagation regime is apparent around $E_{\rm obs}\sim E_{\rm GZK}$. 
Sub-GZK particles, with long interaction lengths and small gyroradii,
are strongly scattered, while super-GZK particles, with short interaction 
lengths and large gyroradii, are much less affected. 
As expected, the time delay decreases with increasing energy
on average but has a rather wide spread at any given energy. 
For comparison, the rectilinear flight time for the mean separation
of sources, $l_s=$ 40 Mpc, is $t_{rec}\equiv l_s/c \approx 10^{8}$
yr. 

Figure 7 shows the fractions of recorded events in all six
$\Lambda$CDM simulations as functions of $\theta$ and $t_d$,
$df/d\theta$ and $df/d \log t_d$, and their cumulative distributions.
The events are divided into three energy channels as in Figure 5,
and each curve is normalized by the total number of events in
the corresponding channel.
In order to demonstrate the dispersion of the statistics due to cosmic 
variance, we also plot the error bars in the cumulative distributions, 
which are calculated as the standard deviations of the values of $f$
for the six simulations.

In the lowest energy channel ($10~{\rm EeV}<E_{\rm obs}<30$ EeV, {\it red lines}), 
the deflection angle is quite large, with about 70\% of the events arriving
with $\theta > 30^{\circ}$, that is, $f(>30^{\circ}) \approx 0.7$.
Moreover, with $f(>10^8\ {\rm yr}) \approx 0.7$ the time delay 
is much longer than the typical lifetimes of AGNs ($\tau_{AGN} = 0.01-0.1$
Gyr).
In the highest energy channel ($E_{\rm obs} \ge 60$ EeV, {\it black lines}),
on the other hand, about 70 \% of the recorded events arrive with a 
deflection angle smaller than $\sim 15^{\circ}$ and a time delay
less than $\sim 10^7$ yr.
About 35 \% arrive with an angle smaller than $\sim 5^{\circ}$.
This implies that the arrival direction of super-GZK cosmic rays may show 
a positional correlation with the source AGNs and also with the LSS,
and the source AGNs are very likely to still be active for such events.
We note, however, that these results are not restricted to the specific
AGN model.
They can be applied for any UHECR sources that have
a spatial distribution and magnetic field environment similar
to those of luminous X-ray clusters.

We note that the present work produces results different 
from what previous studies have predicted \citep{armen05,dolagetal05}.
Specifically, the deflection angle is smaller than that found by 
Sigl \etal but larger than that of Dolag \etal. 
This should be attributable to the difference in the models for the IGMFs, 
as discussed in \S 2.1. 
We also note that the effects of Galactic magnetic fields are
not included in our analysis.
Recently, \citet{takami07}, for instance, considered several
different models for the Galactic magnetic fields and predicted
that the deflection angle of $10^{19.8}$ eV protons should be greater
than 8$^{\circ}$ toward the Galactic center, while being mostly $3-5^{\circ}$ 
outside a circular region of 30$^{\circ}$ radius around the
Galactic center.

\subsection{Predicted Energy Spectrum}

Here we present the energy flux, $J(E_{\rm obs})$, of the recorded
cosmic-ray events in our propagation experiment.
By applying a weighting factor proportional to $E_{\rm inj}^{-\gamma}$
to each recorded particle, the energy spectra for different values of
$\gamma$ can be constructed. 
The predicted spectra are calculated for injection spectra 
with $\gamma$-values of 2.0, 2.4, and 2.7 and for $D_{\rm min}=3$ and 
$D_{\rm min}=10$ Mpc.
Again, all data from the six $\Lambda$CDM simulations are combined.
Figure 8 shows the resulting spectra, $J(E)$, along with the data
observed at AGASA \citep{nagano00}, HiRes-I \citep{bgg06}, HiRes-II
\citep{zech04}, and Auger \citep{parizot07}.
Since the amplitude of the injection spectrum is not specified,
the amplitude of the predicted $J(E)$ is arbitrary in our model.
Therefore each curve was adjusted by eye to fit the HiRes data below
$E_{\rm GZK}$.

The presence of GZK suppression above 60 EeV is obvious in all the
predicted spectra and the observed data except for the AGASA data.
The predicted spectra for $\gamma=2.4-2.7$ are all consistent with 
the observed data, again with the exception of the AGASA data. 

As shown in the bottom panel of Figure 8, our work predicts that
above $\sim 100$ EeV the flux is much higher with $D_{\rm min}=3$
Mpc than with $D_{\rm min}=10$ Mpc, indicating that the contribution
from nearby sources is important.
Thus, if the injection spectrum has a power-law distribution
extending well beyond the GZK energy as we assume here, 
the implication is that the Auger
experiment, which has Centaurus A in its field of view, may see
a higher flux of super-GZK cosmic rays, compared with experiments in
the Northern Hemisphere such as HiRes and the Telescope Array \citep{fm07}. 
However, it is quite possible that the injection spectrum is limited
to a maximum energy $E_{\rm max}$ set by the age and size
of the astrophysical accelerators or by the GZK energy loss at the acceleration
sites. Moreover, the value of $E_{\rm max}$ may vary with the properties
of the accelerator, rather than maintaining a constant value of $10^3$ EeV. 
In order to settle this issue, 
much better statistics for the energy spectrum and
the arrival directions above 100 EeV are needed. 

\section{Summary and Discussion}

In the search for the astrophysical sources of UHECRs,
it is important to understand how the propagation of these
charged particles is affected by intergalactic magnetic fields 
in the large-scale structure of the universe.
On the other hand, the information imprinted on
the distribution of the UHECR arrival directions may help us
to understand the nature of the IGMFs and their roles in 
the formation and evolution of the LSS and constituent galaxies.  
Considering the limitations of current observational techniques in 
measuring the IGMFs in very low density regions such as filaments and
voids, it is crucial to construct a physically motivated model
to estimate the IGMFs in the LSS. 

In this study, we adopted a new model based on the turbulence dynamo
\citep{ryuetal08} to predict the strength of the IGMFs.
The magnetic field energy is estimated from the local vorticity and
turbulent kinetic energy of flow motions in cosmological simulations
of the LSS formation in a concordance $\Lambda$CDM universe.
For the direction of the IGMFs, the topology of the passive magnetic
fields followed in the cosmological simulations is used.
This approach provides an IGMF model that is independent
of the initial seed fields and does not require any renormalization
to yield the observed field strength in the intracluster medium. 
We predict highly structured IGMFs with characteristic field 
strengths on the order of $10^{-6}$ G in clusters of galaxies and 
$10^{-8}$ G in filaments.
The fields should be much weaker in sheets and voids.

Protons with 10 EeV $\le E_{\rm inj}\le 10^3$ EeV are injected at the
locations of luminous X-ray clusters with $kT>1$ keV.
These sources may represent a population of AGNs 
residing inside host clusters.
This X-ray temperature criterion naturally places the sources at strongly
magnetized regions with $ B_s \sim 0.1 \mu$G with a comoving density 
of $2-3 \times 10^{-5}(\Mpc)^{-3}$.
Then the propagation of the UHE protons is followed through the
structured IGMFs, including the energy losses due to interactions
with the cosmic background radiation 
The UHE protons are recorded at the positions of mock observers located in groups
of galaxies, with halo temperatures in the range 
$0.05 {\rm keV} < kT < 0.5 {\rm keV}$.

Below the GZK energy, the UHE protons come from sources as distant as
$\sim 1$ Gpc.
They are significantly scattered by the IGMFs, resulting in a wide
range of deflection angles, up to 180$^{\circ}$, and have time
delays ranging from $10^7$ to $10^{9.5}$ yr.
On the other hand, the protons above 60 EeV come mostly
from sources within $\sim 100$ Mpc.
About 70\% of them avoid strong deflection and arrive at the observers 
within $\sim 15^{\circ}$ of their source position on the sky with
a time delay of less than $\sim 10^7$ yr. 
About 35\% arrive within $\sim 5^{\circ}$.
This implies that there may exist a correlation between the arrival
direction of super-GZK cosmic rays and the sky position of the corresponding AGNs.
We thus conclude that in the present scenario, UHECR astronomy may be
possible at $E>$ 60 EeV.
Our prediction seems to be consistent with a recent report by the
Auger Collaboration \citep{augerScience} in which the
arrival directions of cosmic rays above 60 EeV in their data were found to be
correlated with the sky position of AGNs within 75 Mpc. 

For any cosmological sources, we expect to see GZK suppression
in the energy spectrum of UHECRs if the injection spectrum 
has a power-law distribution and extends well beyond the
GZK energy.
In this case, nearby sources, within $10 - 20$ Mpc, are expected to
make a significant contribution to the flux above $\sim 100$ EeV.

Finally, as recently reported by Auger \citep{unger07}, some
UHECRs might be heavy nuclei.
In the cosmological-shock model, for example, protons can be
accelerated up to a few times 10 EeV, so heavy nuclei should
dominate the particle flux above that energy 
\citep{krb97,inoue07}.
In a future study, we will consider the propagation of heavy nuclei
from cosmological sources in our model IGMFs, taking into account
photo disintegration, photo pair production, and photo pion production
processes.
We expect that the propagation of UHE iron nuclei ($Z=26$), at least, will
be in the diffusive transport regime as a results of their much smaller gyroradius
($r_g\propto E/Z$) \citep{armen05}.
For intermediate-mass nuclei such as He, C, N, and O, detailed propagation
simulations including secondary particles produced by
photo disintegrations are necessary in order to determine whether astronomy with
UHE nuclei is possible or not.

\acknowledgements
The work of H. K. and S. D. is supported by Korea Science and Engineering Foundation 
through the Astrophysical
Research Center for the Structure and Evolution of the Cosmos (ARCSEC).
The work of D. R. and J. C. is supported by Korea Research Foundation grants
funded by the Korean Government (MOEHRD) (KRF-2007-341-C00020 and
KRF-2006-331-C00136, respectively).
This work was also supported by the Korea Foundation for International Cooperation of
Science and Technology through
the grant K20702020016-07E0200-01610.

\clearpage

\begin{deluxetable}{cccc}
\tablecolumns{3}
\tablewidth{0pc}
\tablecaption{Numbers of Sources and  Observers}
\tablehead{
\colhead{Simulation} & \colhead{Sources} & \colhead{Observers}}
\startdata
$\Lambda$CDM1 & 18  &  1000  \\
$\Lambda$CDM2 & 31  &  1344  \\
$\Lambda$CDM3 & 20  &  1379  \\
$\Lambda$CDM4 & 29  &  1336  \\
$\Lambda$CDM5 & 24  &  1343  \\
$\Lambda$CDM6 & 28  &  1365  \\
\enddata
\end{deluxetable}

\clearpage
\begin{figure}
\vspace{-12cm}
\centerline{\includegraphics[width=18cm]{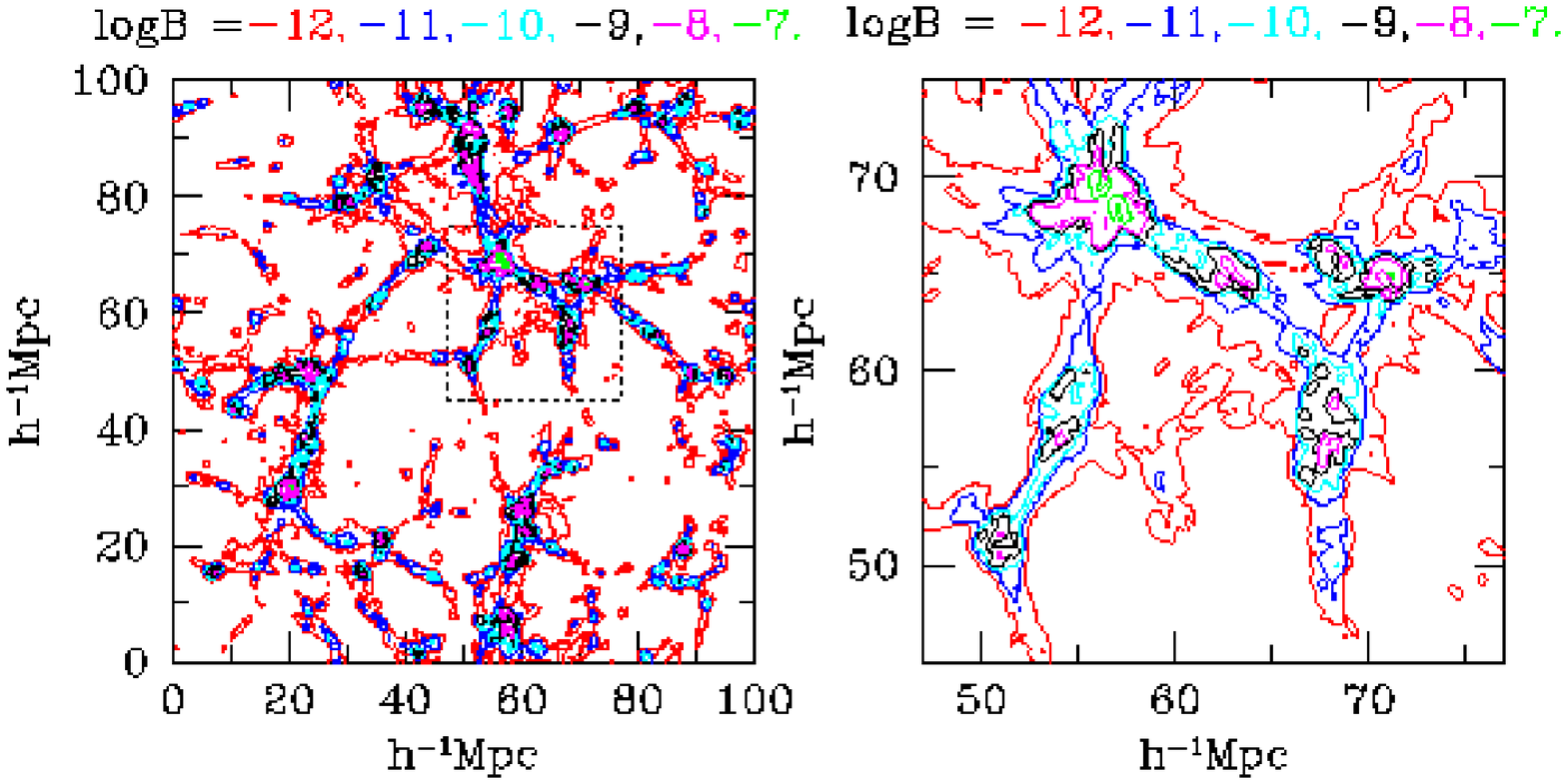}}
\vspace{-4cm}
\figcaption{ {\it Left}:
Two-dimensional slice of area $(100 \Mpc)^2$
showing the distribution of magnetic field 
strength in our model at redshift $z=0$. 
{\it Right}:
Blown-up image of the box delineated with dotted lines in the left panel.
The contour levels are color-coded as follows:
$\log B=$ $-12$ (red), $-11$ (blue), $-10$ (cyan), $-9$ (black), $-8$
(magenta)  and $-7$ (green), where $B$ is the field strength in units
of gauss. 
\label{fig1}}
\end{figure}

\clearpage
\begin{figure}
\vspace{-13cm}
\centerline{\includegraphics[width=18cm]{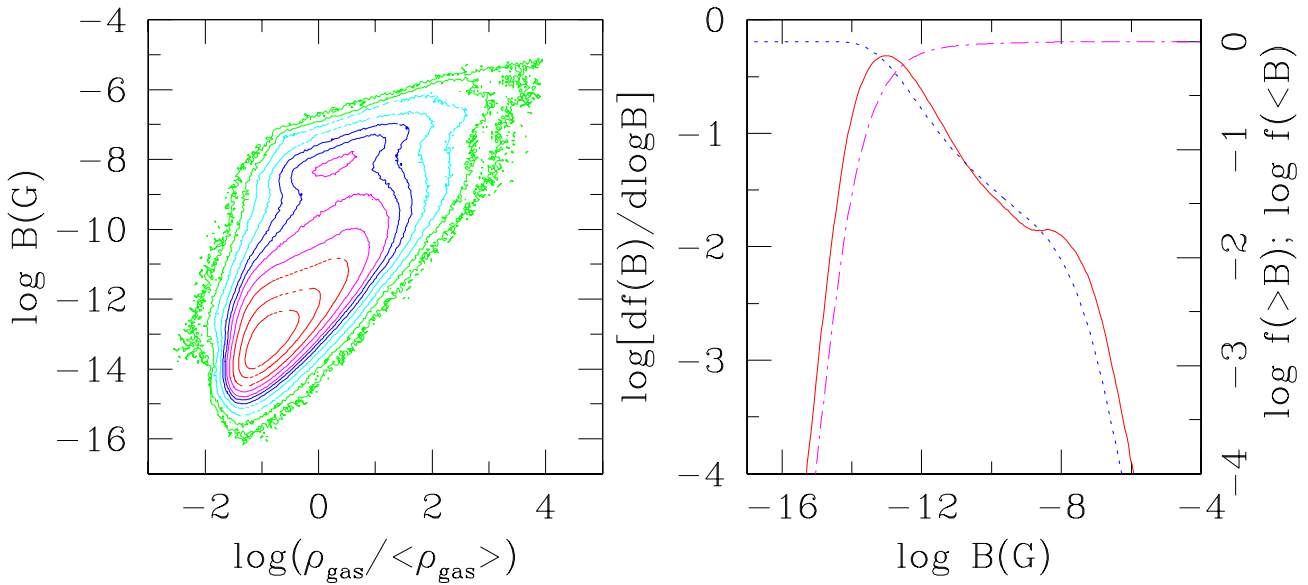}}
\vspace{-4cm}
\figcaption{
{\it Left}: volume fraction in the gas density vs. field strength plane with 
our model IGMFs at redshift $z = 0$. 
{\it Right}: volume fraction, $df/d \log B $ ({\it solid line}),
and its cumulative distributions, $f(>B)$ ({\it dotted line}) and 
$f(<B)$ ({\it dot-dashed line}) as a function of the IGMF strength.
\label{fig2}}
\end{figure}

\clearpage
\begin{figure}
\vspace{-1cm}
\centerline{\includegraphics[width=12cm]{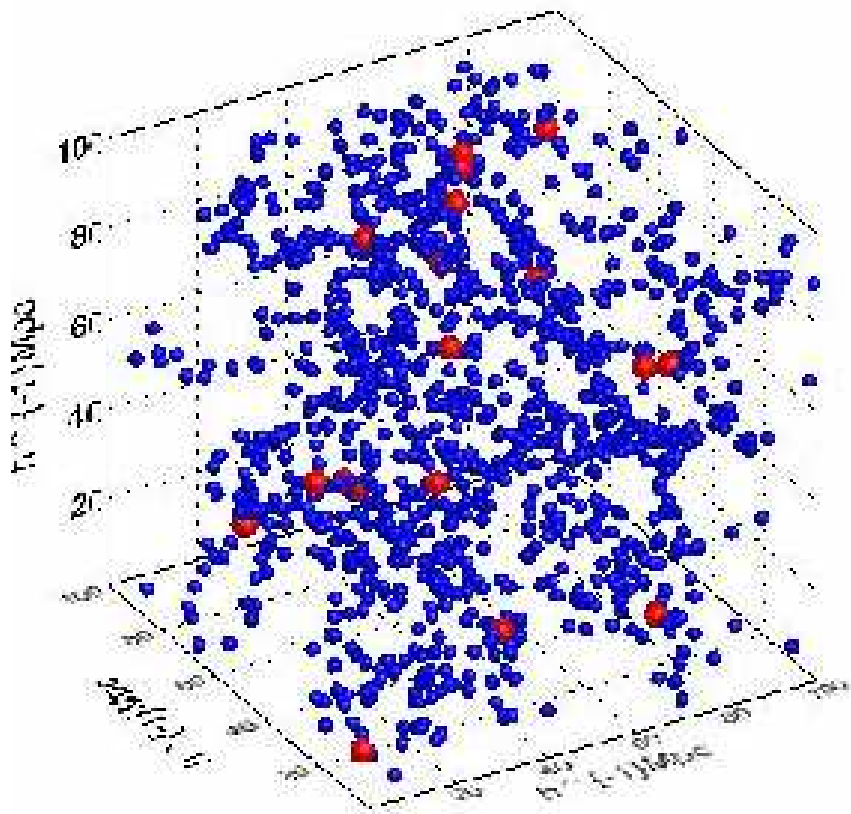}}
\figcaption{
Distribution of UHECR sources ({\it red circles}) and mock observers ({\it blue circles})
in simulation $\Lambda$CDM1.
Sources are modeled as AGNs inside X-ray clusters, while mock observers 
are placed inside groups of galaxies similar to the Local Group.
\label{fig3}}
\end{figure}
 
\clearpage
\begin{figure}
\vspace{-12cm}
\centerline{\includegraphics[width=18cm]{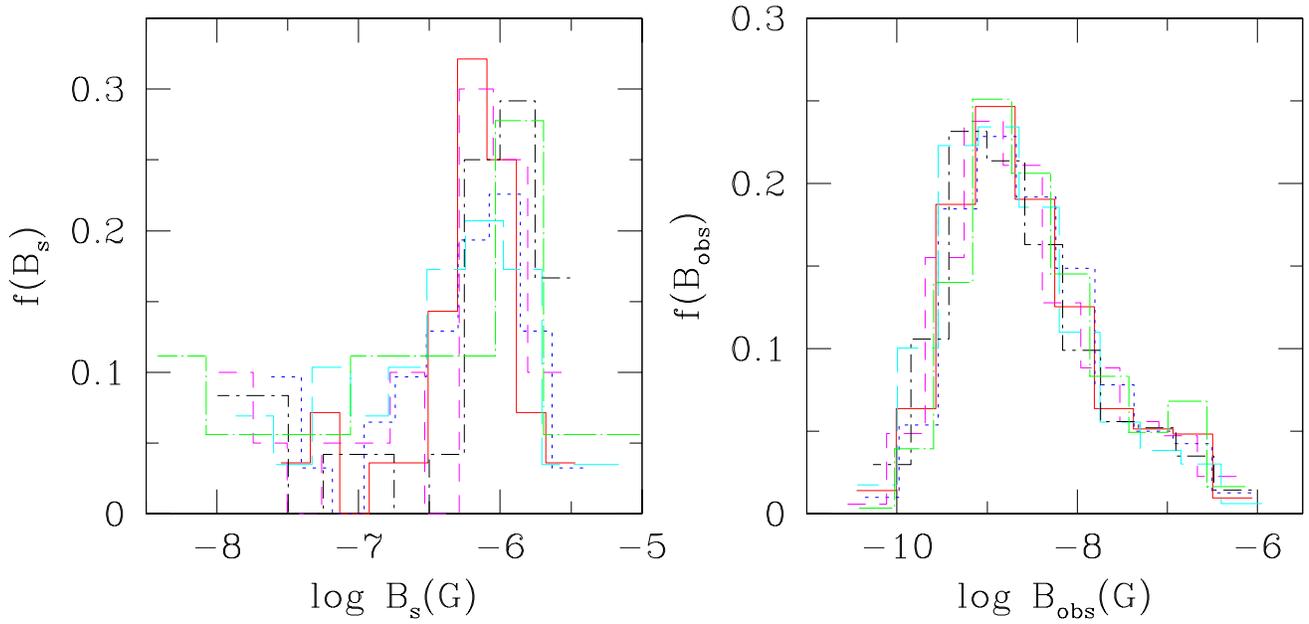}}
\vspace{-4cm}
\figcaption{
{\it Left}: Distribution of magnetic field strength at source
locations in the six simulations with different initial conditions. 
Different line styles are used for simulations $\Lambda$CDM1 - $\Lambda$CDM6.
{\it Right}:
Distribution of magnetic field strength within the observer spheres.
The same line styles as in the left panel are used.
\label{fig4}}
\end{figure}

\clearpage
\begin{figure}
\vspace{-1cm}
\centerline{\includegraphics[width=18cm]{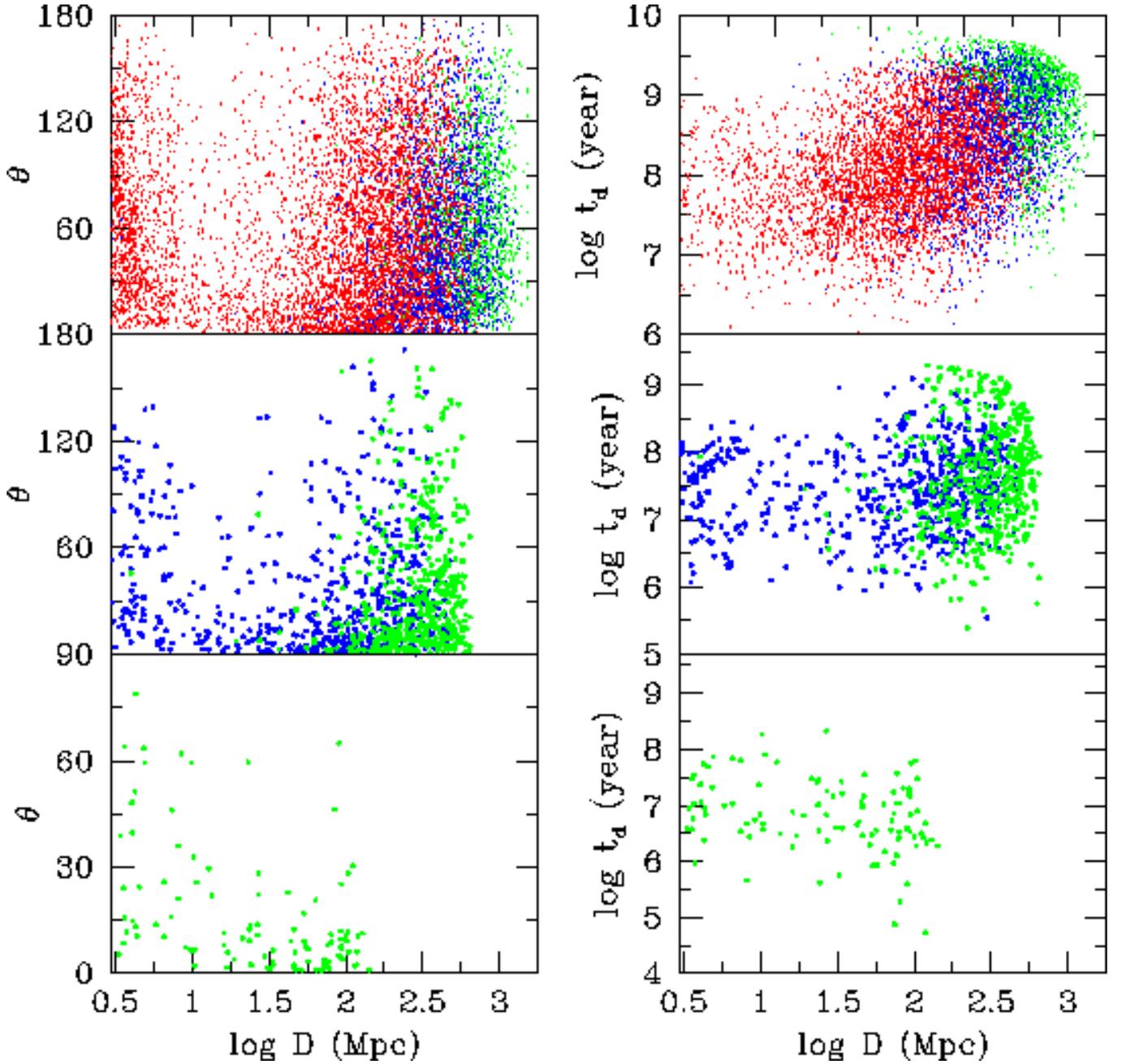}}
\figcaption{
Deflection angle ($\theta$) and time delay ($t_d$) as functions of
the source-to-observer distance $(D)$. 
The events recorded with observed energies $ 10~{\rm EeV}\le E_{\rm obs} < 30$
EeV are shown in the top panels, $ 30~{\rm EeV} \le E_{\rm obs} < 60$ EeV in the
middle panels, and $E_{\rm obs} \ge 60$ EeV in the bottom panels.
The data points are color-coded by the injection energy as follows:
red, $ 10~{\rm EeV} \le E_{\rm inj} < 30$ EeV; blue, $ 30~{\rm EeV} 
\le E_{\rm inj} < 60$ EeV; green, $E_{\rm inj} \ge 60$ EeV. 
An injection spectrum $N(E_{\rm inj}) \propto E_{\rm inj}^{-2.7}$
is assumed.
(See text for details.)
\label{fig5}}
\end{figure}

\clearpage
\begin{figure}
\vspace{-12cm}
\centerline{\includegraphics[width=18cm]{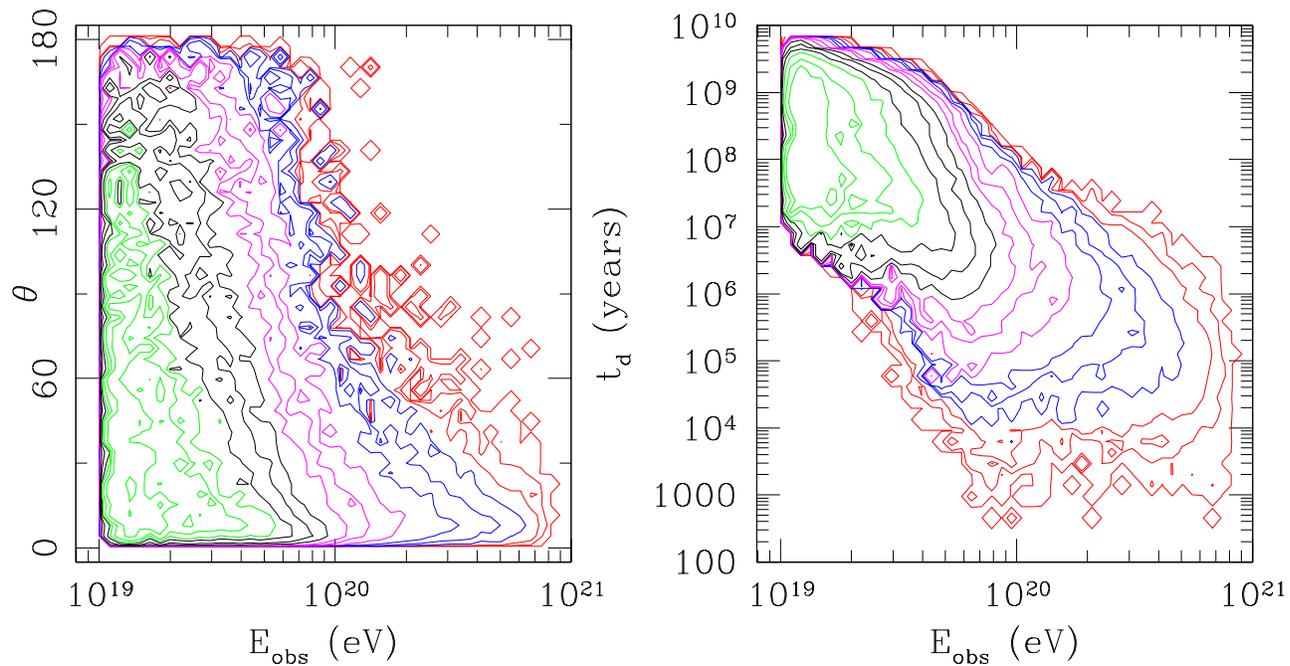}}
\vspace{-4cm}
\figcaption{
Distribution of observed UHECR events in the planes of observed energy
vs. deflection angle ({\it left})  and time delay ({\it right}).
The events recorded in all six $\Lambda$CDM simulations are included,
and $\gamma=2.7$ and $D_{\rm min}=3$ Mpc were used.
\label{fig6}}
\end{figure}

\clearpage
\begin{figure}
\vspace{-1cm}
\centerline{\includegraphics[width=18cm]{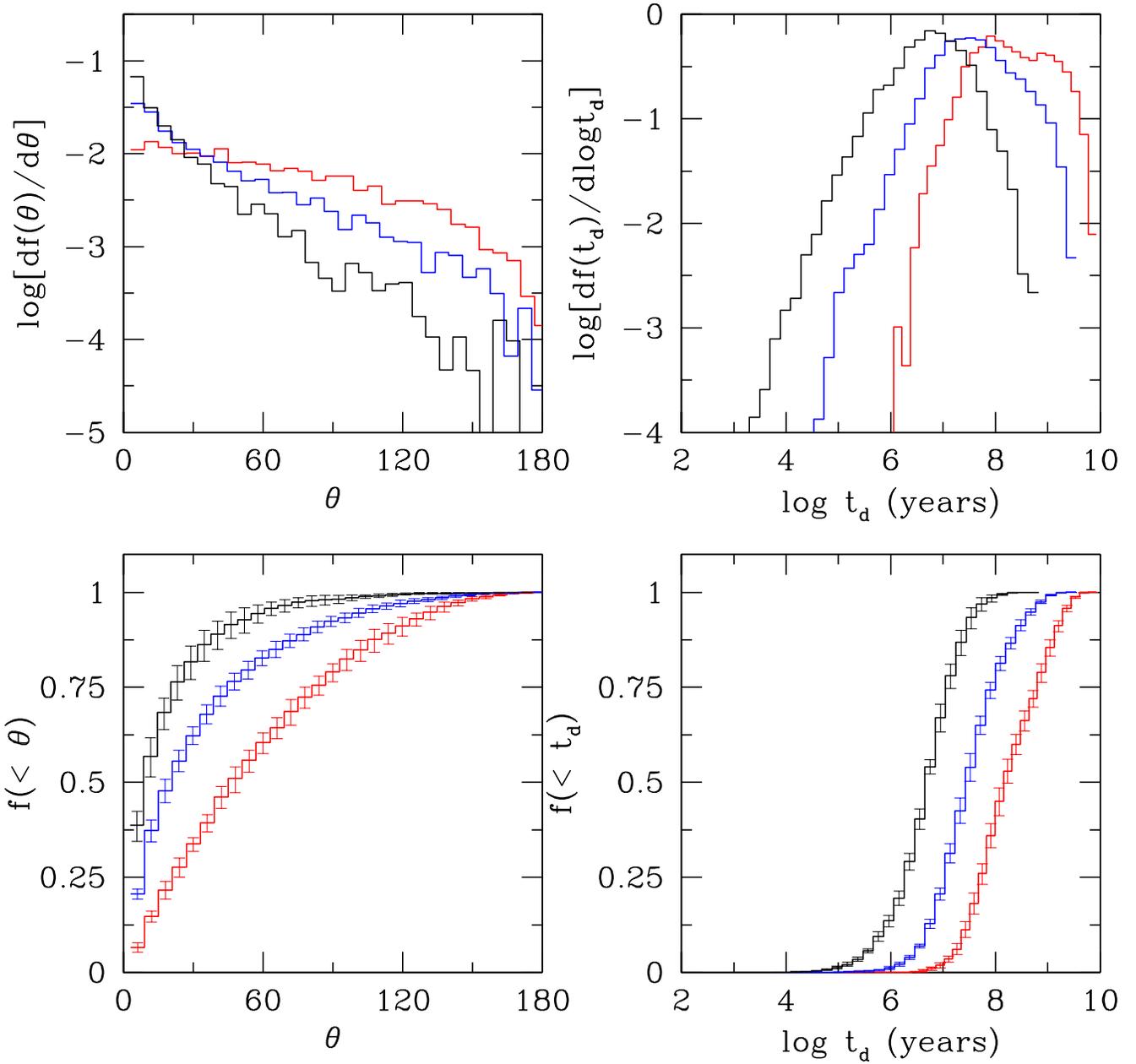}}
\figcaption{
Fractions of observed UHECR events as a function of deflection angle
({\it top left}) and time delay ({\it top right}), and the respective
cumulative distributions ({\it bottom}).
The events recorded in all six $\Lambda$CDM simulations are included,
and $\gamma=2.7$ and $D_{\rm min}=3$ Mpc were used.
The distributions for different energy channels are shown: 
red, $10~{\rm EeV} < E_{\rm obs}< 30$ EeV; 
blue, $30~{\rm EeV} \le E_{\rm obs} < 60$ EeV;
black, $E_{\rm obs}\ge 60$ EeV. 
The error bars shown for the cumulative distributions are the
standard deviations of $f$ for the six simulations.
\label{fig7}}
\end{figure}

\clearpage
\begin{figure}
\vspace{-1cm}
\centerline{\includegraphics[width=18cm]{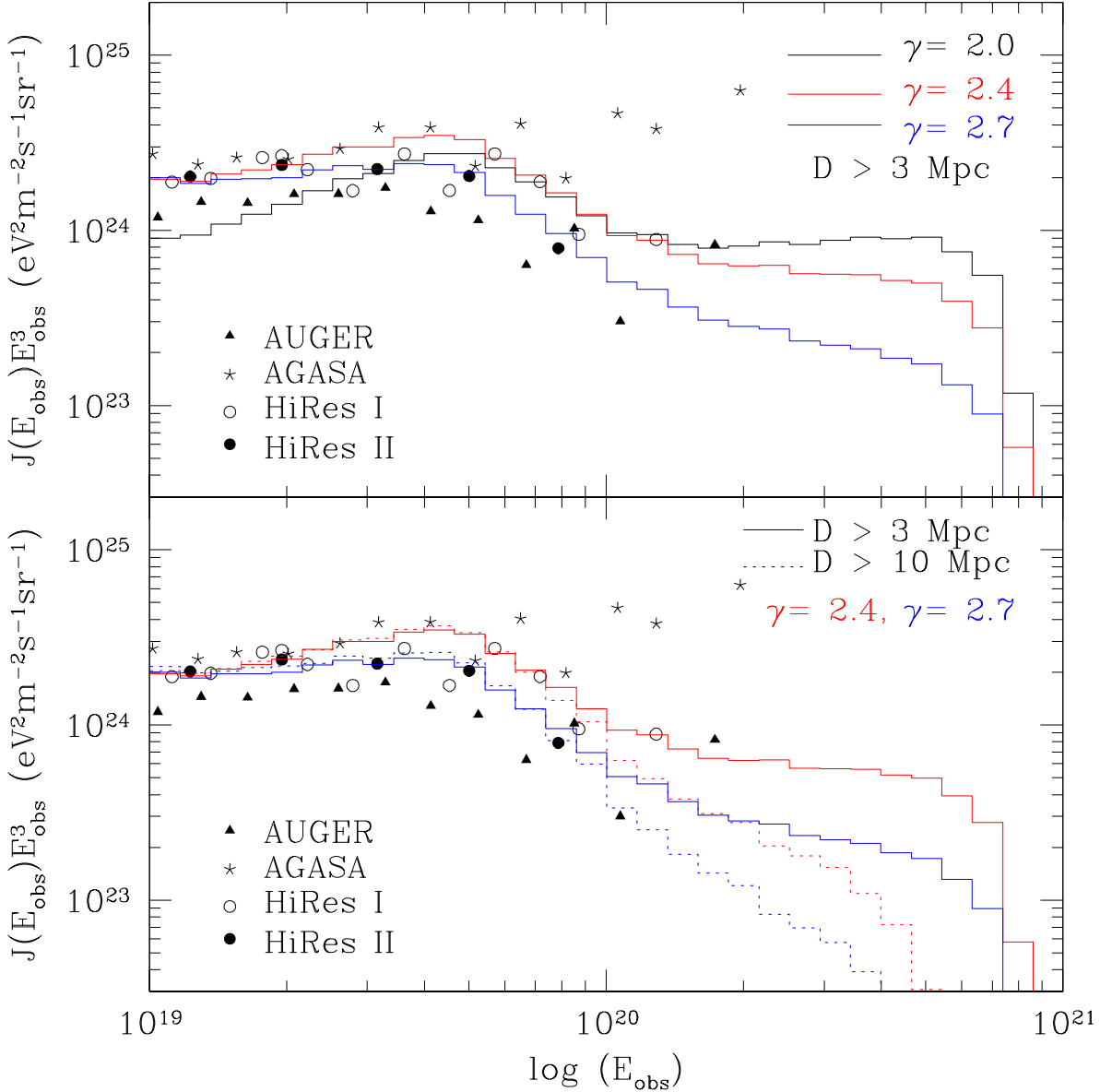}}
\figcaption{
Energy spectra of UHE protons predicted by our model.
The injection spectrum at the sources is proportional to
$E_{\rm inj}^{-\gamma}$ for $10 \le E_{\rm inj} \le 10^3$ EeV. 
{\it Top}: The blue, red, and black lines are for
$\gamma$-values of 2.0, 2.4, and 2.7, respectively, with a minimum
source-to-observer distance $D_{\rm min}=3$ Mpc. 
{\it Bottom}: The red lines are for an injection spectrum
with $\gamma=2.4$, while the blue lines are for $\gamma=2.7$.
The solid lines are for $D_{\rm min}=3$ Mpc, and the dotted lines
are for $D_{\rm min}=10$ Mpc. 
The data observed at AGASA \citep{nagano00}, HiRes-I \citep{bgg06}, 
HiRes-II \citep{zech04}, and Auger \citep{parizot07} are marked
with asterisks, open circles, filled circles, and triangles,
respectively. 
The predicted spectra were arbitrarily scaled by eye to fit the
HiRes data below $E_{GZK}$.
\label{fig8}}
\end{figure}

\end{document}